\documentclass[sigconf, screen, dvipsnames, authorversion]{acmart}

\usepackage{hyperref}
\usepackage[noend]{algorithmic}
\usepackage{algorithm}
\usepackage{multirow}
\usepackage{amsmath}
\usepackage{bbm}
\usepackage{booktabs}
\usepackage[inline]{enumitem} 
\usepackage{subcaption}
\usepackage{bm} 
\usepackage{xcolor}

\usepackage{lipsum}

\usepackage{wrapfig}

\usepackage{arydshln}
\makeatletter
\def\adl@drawiv#1#2#3{
        \hskip.5\tabcolsep
        \xleaders#3{#2.5\@tempdimb #1{1}#2.5\@tempdimb}%
                #2\z@ plus1fil minus1fil\relax
        \hskip.5\tabcolsep}
\newcommand{\cdashlinelr}[1]{%
  \noalign{\vskip\aboverulesep
          \global\let\@dashdrawstore\adl@draw
          \global\let\ adl@draw\adl@drawiv}
  \cdashline{#1}
  \noalign{\global\let\adl@draw\@dashdrawstore
          \vskip\belowrulesep}}
\makeatother

\usepackage{tikz}
\usetikzlibrary{automata, arrows, bayesnet, bending}

\usepackage{tikzscale}

\DeclareMathOperator*{\argmax}{arg\,max}

\copyrightyear{2025}
\acmYear{2025}
\setcopyright{acmlicensed}\acmConference[RecSys '25]{Proceedings of the
Nineteenth ACM Conference on Recommender Systems}{September 22--26,
2025}{Prague, Czech Republic}
\acmBooktitle{Proceedings of the Nineteenth ACM Conference on
Recommender Systems (RecSys '25), September 22--26, 2025, Prague, Czech
Republic}
\acmDOI{10.1145/3705328.3748011}
\acmISBN{979-8-4007-1364-4/2025/09}
\begin{document}
\title{Counterfactual Inference under Thompson Sampling}

\author{Olivier Jeunen}
\affiliation{
  \institution{aampe}
  \city{Antwerp}
  \country{Belgium}
}
\email{olivier@aampe.com}

\begin{abstract}
Recommender systems exemplify sequential decision-making under uncertainty, strategically deciding what content to serve to users, to optimise a range of potential objectives.
To balance the explore-exploit trade-off successfully, Thompson sampling provides a natural and widespread paradigm to probabilistically select which action to take.
Questions of causal and counterfactual inference, which underpin use-cases like offline evaluation, are not straightforward to answer in these contexts.
Specifically, whilst most existing estimators rely on action propensities, these are not readily available under Thompson sampling procedures.

We derive exact and efficiently computable expressions for action propensities under a variety of parameter and outcome distributions, enabling the use of off-policy estimators in Thompson sampling scenarios.
This opens up a range of practical use-cases where counterfactual inference is crucial, including unbiased offline evaluation of recommender systems, as well as general applications of causal inference in online advertising, personalisation, and beyond.\end{abstract}

\maketitle

\section{Introduction \& Motivation}
Recommender systems govern content consumption on the world wide web, making automated decisions about which content to surface to which users, billions of times per day across a broad spectrum of applications and use-cases.
Sequential algorithmic decision-making under uncertainty is a decades-old research topic that has spawned a vast literature~\cite{robbins1952some}, recently often referred to as (contextual) \emph{bandit} methods~\cite{lattimore2020bandit,vandenAkker2024}.
At a high level of abstraction: a decision-maker iteratively observes some context $X$ and needs to decide on an action $A$ to take.
The decision-maker observes a reward $R$, and aims to take actions that maximise the cumulative rewards they will obtain over some time horizon.
In a recommendation use-case, the context might include a user identifier among other relevant features, the action could be an item (or a set or ranking of items), and the rewards could be various types of interactions.

Thompson sampling (TS) is a popular heuristic for probabilistic action selection~\cite{Thompson1933, Russo2018}, providing an intuitive solution with near- or asymptotically optimal regret~\cite{Kaufmann2012,Jin2023}, as well as strong empirical performance~\cite{Chapelle2011}.
As a result, it has become a widespread method for various practical applications that make use of bandit algorithms, including recommender systems~\cite{Broden2018, Lu2018, Bendada2020, Jeunen2021_TS, Eide2022, Su2023, Zhu2023,Su2024}.

TS leverages a posterior distribution over the parameters of a reward model, to select the action that maximises the estimated reward according to a randomly drawn belief about said model.

When framing the recommendation task as a decision-making problem~\cite{CONSEQUENCES2022, Jeunen2023}, it is natural to consider applications of causal or counterfactual inference~\cite{Wang2020_Causal,Gao2024,Bottou2013,Saito2021,Jeunen2021Thesis}.
Methods from these adjacent fields can be used to answer general aggregate-level questions of the form: ``\emph{What would the distribution of $R$ have been, had we deployed this alternative decision-making policy instead}?''~\cite{Vasile2020}

The most common embodiment of such questions arises when carrying out offline evaluation tasks, aimed at estimating (differences in~\cite{Jeunen2024_DeltaOPE}) reward metrics for potential alternative recommendation models~\cite{Yang2018,Jeunen2019,Gilotte2018,Gruson2019,Jeunen2023_nDCG}.
Additional to performance evaluation, these methods open up avenues for \emph{learning} and \emph{fairness} in recommendation---see~\citet{Joachims2021} for a broad overview.

Importance sampling or Inverse Propensity Scoring (IPS)~\cite[\S 9]{Owen2013} is the foundation that powers virtually all counterfactual estimators.
At its core, it requires accurate action probabilities (often called \textit{propensities}) to address the mismatch between the data-collecting \textit{logging} policy, and the so-called \textit{target} policy. 
In TS settings, where the randomisation over actions is a reflection of the uncertainty in the reward model, rather than an explicit distribution over actions themselves, this is not a straightforward quantity to estimate.
\citet{Amat2018} explicitly list the lack of an efficient and effective method to compute propensities as the main downside of using TS, as it requires them to fall back to the Replay method for offline evaluation~\cite{Li2011} which requires a uniformly random logging policy. 
As a result, it is significantly more costly and less sample-efficient than more recent alternatives that rely on IPS~\cite{Swaminathan2015,Dudik2014,Su2020,Saito2022_MIPS, Saito2023, Gupta2024}.

Our work aims to close this gap.
We derive a general expression for action propensities under TS, reducing to an integral that involves the probability density functions (PDFs) and cumulative distribution functions (CDFs) for the action-specific reward distributions.
For the most common settings of (log)normal distributions for continuous outcomes and Beta distributions for binary outcomes, we show that this integral reduces to an efficiently computable multivariate normal CDF or an analytically computable expression, respectively.
We show that, under a common linearity assumption on the last layer of the reward model~\cite{Chapelle2011,Watson2021}, a normal distribution on model parameters directly translates to a distribution over model outcomes (i.e. rewards)~\cite{Zhang2021}, enabling efficient propensity calculations.
These novel derivations open up a range of potential use-cases where either or both the logging and target policies make use of TS or related action-selection procedures.

We empirically validate our theoretical contributions by showcasing the efficacy of offline evaluation methods that would not have been feasible before, through reproducible experiments leveraging the Open Bandit Pipeline (OBP)~\cite{Saito2021_OBP}.
\section{Background \& Related Work}
\subsection{Counterfactual or Off-Policy Estimation}
It is common to have access to data collected under a logging policy $\pi_{0}$ (e.g. a deployed recommender system), and want to answer questions about an alternative policy $\pi_t$ (e.g. a new model we want to test).
This is often referred to as \emph{off-policy estimation}.
Existing methods typically rely on some form of importance sampling or Inverse Propensity Scoring (IPS)~\cite[\S 9]{Owen2013}.
In a standard contextual bandit setup with contexts $X$, actions $A$, rewards $R$, and policy parameters $\Theta$; IPS can be used to obtain counterfactual reward estimates for varying instantiations of $\Theta$.
With an implied expectation over contexts $\mathop{\mathbb{E}}_{x \sim \mathsf{P}(X)}$ and $\pi_{i}(a|x)\equiv\mathsf{P}(A=a|X=x;\Theta=\theta_i) $ we have:
\begin{equation}\label{eq:IPS_expectation}
    \mathop{\mathbb{E}}_{a \sim \pi_{t}(\cdot|x)}\left[R\right] =
    \mathop{\mathbb{E}}_{a \sim \pi_{0}(\cdot|x)}\left[R\frac{\pi_{t}(a|x)}{\pi_{0}(a|x)}\right].
\end{equation}
That is, we can unbiasedly estimate the expected reward under counterfactual parameter values $\theta_{t}$ (LHS) using data collected under logging parameter values $\theta_{0}$ (RHS).
Naturally, we require a ``common support'' assumption to ensure that the probability ratio cannot imply a division by zero.
\citet{Chandak2021} show that with a simple adaptation, we can model the full cumulative distribution of $R$, along with valid high-confidence bounds. 
We refer the interested reader to their work, but highlight that the core idea still leverages IPS, and hence, access to propensities $\pi(a|x)$ also enables theirs and similar methods to be applied.

Given a dataset $\mathcal{D} = \{(x_i,a_i,r_i)_{i=1}^{m}\}$ collected under $\pi_{0}$, Eq.~\ref{eq:IPS_expectation} translates into an empirical IPS estimate for the average reward as:
\begin{equation}
    \widehat{R}_{\rm IPS}(\mathcal{D},\pi_t) = \frac{1}{|\mathcal{D}|} \sum_{(x,a,r) \in \mathcal{D}} r\frac{\pi_t(a|x)}{\pi_0(a|x)}.
\end{equation}
Whilst this estimator is unbiased, its variance is often problematic, and a large literature on off-policy estimation has focused on variance reduction.
Some trade in variance for a small bias~\cite{Ionides2008,Swaminathan2015}, others leverage additive control variates~\cite{Gupta2024,Dudik2014,Vlassis2019, Farajtabar2018, Su2020,Jeunen2024_DeltaOPE}.

Several related works have leveraged variants of these estimators to achieve empirical success in recommender systems evaluation~\cite{Yang2018, Gilotte2018,Gruson2019,Jeunen2023_nDCG,Jeunen2023_C3PO}, as well as learning~\cite{Jeunen2024Multi,Jeunen2021_Pessimism,Jeunen2020,chen2019top,ma2020off}.

In the adjacent field of Information Retrieval and the task of Learning-to-Rank (LTR), \emph{unbiased} LTR has recently seen a surge of interest~\cite{Joachims2017, Oosterhuis2020, Gupta2024_CIKM, Gupta2024_WSDM}; and these methods have also seen adoption in various areas of computational advertising~\cite{Bottou2013, Jeunen2023_AuctionGym, Sagtani2024}. 

In practice, the policy that decides which actions to take in which contexts is often parameterised \emph{explicitly} as a probability distribution over $A$.
Imagine $\theta$ representing the mean and covariance matrix for a Gaussian distribution $\mathcal{N}(\mu;\Sigma)$, or the parameters for a neural network with a softmax output.
This modelling choice permits direct application of IPS-based estimators, but complicates several other desiderata.
It remains, for example, unclear how to balance exploration with exploitation in a principled manner.
Additionally, accurate reward estimates for $\mathsf{P}(R|A;X)$ are often part of a larger eco-system, as they inform e.g. features for other models, or serve as input for real-time bidding in advertising applications.

The consequence is that alternative modelling paradigms from the broader literature on sequential decision-making under uncertainty remain popular to inform practical scenarios.

\subsection{Thompson Sampling}
A common heuristic algorithm is Thompson sampling (TS), focusing on the \emph{value} of individual actions to make decisions. 

TS leverages a posterior distribution on the parameters $\theta$ of a reward model $f_{\theta}$ to probabilistically select actions to take.
That is, we sample model parameters and subsequently compute reward estimates according to these sampled parameters, selecting the action with the maximal estimate as:
\begin{equation}
a^{\star} =\argmax_{a \in \mathcal{A}} f_{\tilde \theta}(a,x),\qquad \text{~where~}\tilde\theta\sim \widehat{\mathsf{P}}(\Theta|\mathcal{D}).
\end{equation}
The estimated posterior distribution $\widehat{\mathsf{P}}(\Theta|\mathcal{D})$ can be updated instantaneously when new data arrives, but this is typically done in batches of delayed feedback over time~\cite{Chapelle2011,Jeunen2021_TS}.
TS is attractive to practitioners as it provides an intuitive and effective way to deal with the explore-exploit trade-off that permeates sequential decision-making problems in practice.
Indeed, variants of TS have been shown capable of attaining minimax optimal regret~\cite{Jin2023}.

In the case of binary outcomes, a Beta-Bernoulli model is often appropriate, where the parameter $\theta$ would correspond to the probability of success in a Bernoulli($\theta$) distribution, and $\mathsf{P}(\theta|\mathcal{D})$ is a Beta distribution~\cite[\S 4.6.2]{Murphy2022}.
Alternatively, when a general machine learning model $f_{\theta}$ is used to estimate the reward for a context-action pair $\hat r(a,x)\approx f_{\theta}(a,x)$, the posterior over $\theta$ is typically modelled with a (multivariate) normal distribution~\cite{Chapelle2011}.
In modern applications where $f_{\theta}$ can be a high-dimensional neural network, the dimensionality of the parameter space can render the posterior arduous to maintain.
As such, \citet{Zhang2021} reframe TS with a posterior on the reward instead, showing that this matches the regret of existing modelling alternatives.
\section{Methodology \& Contributions}
Our goal is to derive efficiently computable expressions for action propensities under TS regimes, for a variety of parameter and outcome distributions.
Following the framework of \citet{Zhang2021} with a posterior distribution on the reward, we have a set of estimated reward distributions $f_{\theta}(a,x) \forall a \in \mathcal{A}$ and wish to estimate:
\begin{align}
\pi_{\theta}(a_i|x) &=
\mathsf{P}_{\max}(a_i|\mathcal{A},\Theta=\theta,X=x) \nonumber\\
&\equiv \mathsf{P}(i=\argmax_{a \in \mathcal{A}}\{\hat{r}_{a} \sim f_{\theta}(a,x)\}).    
\end{align}
When parameters are clear and unambiguous from context, we will refer to this quantity as $\mathsf{P}_{\max}(a_i|\mathcal{A})$.

Alternatively, with a posterior distribution on the model parameters $\mathsf{P}(\Theta|\mathcal{D})$, we will show in Section~\ref{sec:param_uncertainty} that this can be equivalently rewritten to a distribution on outcomes, when we consider a model class where the parameters have a linear relationship to the outcome---including Bayesian generalised linear models~\cite{Chapelle2011}, or neural networks with uncertainty on the final layer~\cite{Watson2021,Su2024}.

Denote with $f_i \equiv f_{\theta}(a_i, x)$  the PDF of the estimated reward distribution, and with $F_i$ its CDF. 
For notational convenience and w.l.o.g. we will assume that we are interested in the propensity of action $a_1$, as actions can simply be relabelled to ensure this. Furthermore, $|\mathcal{A}|\equiv n$.
Assuming a continuous reward distribution, we can write:
\begin{align}
    \pi_{\theta}(a_1|x) &= \int\limits_{-\infty}^{+\infty}\underbrace{\int\limits_{-\infty}^{r_1}\ldots\int\limits_{\infty}^{r_1}}_{\forall j \in [2,n]} f_1(r_1)f_2(r_2) \ldots f_n(r_n){\rm d}r_1{\rm d}r_2\ldots{\rm d}r_n \nonumber \\
    &= \int\limits_{-\infty}^{+\infty} f_1(r_1) \left(\prod_{j \in  [2,n]} F_j(r_1) \right){\rm d} r_{1}.
    \label{eq:propensity_integral}
\end{align}

Intuitively, for every possible value of $r_1$, the reward estimates for all other actions need to be lower for $a_1$ to be selected. 
In the general case, this integral does not have a simple analytical solution.
In what follows, we derive efficiently computable or closed-form solutions for common families of reward distributions: (log)normal distributions for continuous outcomes, and Beta distributions for binary rewards.

\subsection{Continuous Rewards: Normal Distributions}
For Gaussian densities, a solution to Eq.~\ref{eq:propensity_integral} is known~\cite{Hartmann2017}.
Fixing $x$ and denoting the parameters for reward distributions for individual actions as $\theta=\{(\mu_i,\sigma_i^2) \forall a_i \in \mathcal{A}\}$, we can rewrite the full integral as the output of the CDF of a single multivariate normal distribution: 
\begin{gather}
    \pi_{\theta}(a_1|x)  = F(\bm{\mu}|\bm{m},\bm{V}), \text{~where~} \bm{m} = [\mu_2 \ldots\mu_n]^{\intercal}, \bm{\mu} = \mu_1 \bm{1}_{n-1} \nonumber \\
    \text{and~}
    \bm{V} = \begin{bmatrix}
    \sigma_2^2+\sigma_1^2 & \dots & \sigma_1^2 \\
\vdots & \ddots & \vdots \\
\sigma_1^2 & \dots & \sigma_n^2 + \sigma_1^2 \\
\end{bmatrix}.
\label{eq:gaussian_propensity}
\end{gather}
Note that this multivariate Gaussian CDF is an integral without a known analytical expression in and of itself---but one that is widespread, and where many accurate and efficient implementations are available to evaluate it~\cite{Virtanen2020}.
Similar quantities appear in the literature on PAC-Bayesian policy learning~\cite{Sakhi2023}.

\subsubsection{Uncertainty on Model Parameters}\label{sec:param_uncertainty}
Traditional TS places uncertainty on model \emph{parameters} instead of model \emph{output}.
That is, instead of directly sampling outputs as $\hat{r}_{a} \sim f_{\theta}(a,x)$, we sample model parameters $\tilde\theta \sim \mathsf{P}(\Theta)$ and simply compute reward estimates as $\hat r_a= f_{\tilde \theta}(a,x)$, before picking the maximal one.
It should be noted that in many common practical instantiations, these are equivalent. If we assume the parameters over which to place uncertainty $\theta\in\mathbb{R}^d$ to correspond to weights in a linear model and $f(a,x)$ is a fixed model transformation (e.g. a neural network, or the identity function in the case of a linear model), we have:
\begin{gather}
\mathsf{P}(\theta) = \mathcal{N}(\bm{\mu}_{\theta};\bm{\Sigma}_\theta), \qquad R=\theta^\intercal f(a,x), \nonumber \\
\mathsf{P}(R|A=a,X=x) = \mathcal{N}\left(\bm{\mu}_{\theta}^{\intercal}f(a,x); f(a,x)^{\intercal}\bm{\Sigma}_\theta f(a,x)\right). \qquad\end{gather}
Naturally, these distributions can equivalently be plugged into Eq.~\ref{eq:gaussian_propensity}, and are hence covered by our theoretical and empirical analysis.
Additionally, note that monotonic transformations on model outputs (e.g. the logistic sigmoid function), do not affect which action gets selected and hence do not affect action propensities.

\subsubsection{Log-Normal Distributions}
Another plausible modelling choice in recommendation applications to represent continuous rewards, is to leverage log-normal distributions.
One can imagine non-negative, continuous and typically skewed outcomes like gross merchandise value or dwell time; but log-normal distributions also occur more widely in reinforcement learning value distributions~\cite{Zhang2017}.
When $\bar{R}$ follows a $\mathcal{N}(\mu;\sigma^2)$ distribution, then $R\coloneq e^{\bar{R}}$ follows a $\mathrm{Lognormal}(\mu;\sigma^2)$ distribution.
Because the exponential transformation is monotonic, action propensities can be computed on $\bar{R}$ itself and we can directly apply the derivation for normal distributions here as well.

\subsection{Binary Rewards: Beta Distributions}
When rewards are binary, $\mathsf{P}(R=1|A,X)$ is typically modelled with a Beta distribution.
Conditional on $X=x$, this implies a set of parameters $\theta = \{(\alpha_i,\beta_i) \forall a_i \in \mathcal{A}\}$.
Note that these parameters themselves can be the output of any other learnt model.

To the best of the author's knowledge, an analytically computable solution to Eq.~\ref{eq:propensity_integral} for Beta distributions is not generally available.
In what follows, we first solve a simplified version of the problem to provide the building blocks necessary for a generalised solution.

\paragraph{Two actions.}
In the case of two potential actions, the problem simplifies to estimating the probability that any one Beta distribution is more likely to yield a higher-value sample.
In the context of Bayesian A/B-testing, this problem has received some attention before~\cite{Miller2015}.
For all actions $a_{i} \in \mathcal{A}$, write $p_{i} \sim {\rm Beta}(\alpha_i, \beta_i)$ and assume $\alpha_i,\beta_i \in \mathbb{N}$.\footnote{Note that this holds per definition in a standard Beta-Bernoulli setup where $\alpha$ and $\beta$ count positive and negative rewards respectively.}
For any two actions $(a_{i}, a_{j}) \in \mathcal{A}$, \citet{Miller2015} provides a derivation leveraging the Beta function:
\begin{equation}\label{eq:two_actions}
    \mathsf{P}(p_{i} > p_{j}) =   \sum_{m=0}^{\alpha_i-1}{\frac{B(\alpha_j+m, \beta_i + \beta_j)}{(\beta_i+m)B(1+m,\beta_i)B(\alpha_j,\beta_j)}}.
\end{equation}
Appendix~\ref{app:miller} reproduces \citet{Miller2015}'s derivation for completeness.

Eq.~\ref{eq:two_actions} gives an analytical expression for the probability that action $i$ is preferred over action $j$ by the TS procedure.
\citet{Miller2015} further derives probabilities extended to three and four actions by hand.
In this manuscript, we aim to derive a generalised solution that yields analytically computable propensities under any $n$-sized action space.
As a basis for our solution, we first lay out \citet{Miller2015}'s procedure.

\paragraph{Extending to $n$ actions.}
Considering Eq.~\ref{eq:propensity_integral}, we can plug in the regularised incomplete beta function for the CDFs: 
\begin{align}\label{eq:prod_of_Is}
    \pi_{\theta}(a_1|x)  &= \int\limits_{0}^{1} f_1(r_1) \left(\prod_{j \in  [2,n]}  I_{r_1}(\alpha_j, \beta_j)\right){\rm d} r_{1}.
\end{align}
Leveraging the identities:
\begin{align}
\label{eq:beta_inverse_identity}
I_{r_i}(\alpha, \beta) &= 1 - I_{1-{r_i}}(\beta, \alpha) \nonumber\\
&= 1 - \sum_{k=0}^{\alpha-1} \frac{r_i^{k}(1 - r_i)^\beta}{(\beta+k)B(1+k,\beta)},
\end{align}
we can separate Eq.~\ref{eq:prod_of_Is} into a sum of probability integrals and an unidentified rest term.
Defining a shorthand for all subsets of actions of size $n$ as $\mathcal{A}^{n}_{\setminus i}\equiv\left\{S  \subseteq \mathcal{A}_{\setminus i} \mid~\mid S\mid=n \right\}$, we can express the result of \citet{Miller2015}'s derivation as:
\begin{align}
&\pi_{\theta}(a_1|x)  = \nonumber\\\label{eq:incl-excl2}
&1 + \sum_{i=1}^{n-2} (-1)^{i}\left(\sum_{S \in \mathcal{A}^{i}_{\setminus 1}} \mathsf{P}_{\min}(a_1|S\cup \{a_1\})\right) \\
+ &(-1)^{n-1}\sum_{j_{2}=0}^{\alpha_{2}-1} \ldots\sum_{j_{n}=0}^{\alpha_{n}-1}\frac{B\left(\alpha_{1}+\sum\limits_{a=2}^{n}j_{a} , \sum\limits_{a=1}^{n}\beta_{a}\right)}{B(\alpha_1,\beta_1)\prod\limits_{a=2}^{n}(\beta_a+j_a )B(1+j_a,\beta_a)}.\nonumber
\end{align}

The probability $\mathsf{P}_{\min}(a_1 |S\cup \{a_1\})$ for sets of size $n$ is then derived by hand.
Intuitively, we iterate over all subsets of actions in $\mathcal{A}_{\setminus 1}$ that yield sampled beliefs that are all higher than $r_1$ and sum their probabilities, with a remainder term.
This provides a closed-form expression for our estimand, following the mathematical tools \citet{Miller2015} leveraged in their derivation.
In what follows, we shed further light on this existing formula and provide a simplification.

\paragraph{The Inclusion-Exclusion Principle and Symmetry}
The probability that $r_1$ is maximal can alternatively be expressed as a conjunction of pairwise comparisons.
Iteratively applying the Inclusion-Exclusion Principle~\cite[\S 1.6.9]{Durret2019}, we then obtain:
\begin{align}
    \mathsf{P}_{\max}(a_1| \mathcal{A}) &= \mathsf{P}\left(\bigcap_{j = 2}^{n} p_1>p_j\right) \nonumber\\
    &= 1 + \sum_{i=1}^{n-1} (-1)^{i}\left(\sum_{S\in\mathcal{A}^{i}_{\setminus 1}} \mathsf{P}_{\min}\left(a_1 |S\cup \{a_1\}\right)\right).\label{eq:incl-excl3}
\end{align}
Contrasting Eqs.~\ref{eq:incl-excl3} and~\ref{eq:incl-excl2}, we can identify the remainder term from \citet{Miller2015} as:
\begin{gather}
    \mathsf{P}_{\min}(a_1|\mathcal{A}) = \sum_{j_{2}=0}^{\alpha_{2}-1} \ldots\sum_{j_{n}=0}^{\alpha_{n}-1} \frac{B\left(\alpha_{1}+\sum\limits_{a=2}^{n}j_{a} , \sum\limits_{a =1}^{n}\beta_{a}\right)}{B(\alpha_1,\beta_1)\prod\limits_{a=2}^{n}(\beta_a+j_a )B(1+j_a,\beta_a)}.\label{eq:lowest_of_set}
\end{gather}
This expression provides a generalised formula for the probability that an action will have a randomly drawn belief that is \textit{minimal} over the set of all actions.

On our quest to find the probability that action $i$ is \emph{maximal}, we have now identified the probability that it is \emph{minimal}.
Intuitively, this problem should be equally complex.
This intuition can easily be verified by a symmetry argument on the Beta distribution.
Let $p_i\sim {\rm Beta}(\alpha_i,\beta_i), p_j\sim {\rm Beta}(\alpha_j,\beta_j), \bar{p_i}\sim {\rm Beta}(\beta_i,\alpha_i), \bar{p_j}\sim {\rm Beta}(\beta_j,\alpha_j)$:
\begin{equation}
\mathsf{P}(p_i>p_j) = \mathsf{P}(\bar{p_i} < \bar{p_j}).
\end{equation}
As a direct consequence, we have that:
\begin{gather}
    \mathsf{P}_{\max}(a_1|\mathcal{A}) =\sum_{j_{2}=0}^{\beta_{2}-1} \ldots\sum_{j_{n}=0}^{\beta_{n}-1}
\frac{B\left(\beta_{1}+\sum\limits_{a=2}^{n}j_{a} , \sum_{a=1}^{n}\alpha_{a}\right)}{B(\beta_1,\alpha_1)\prod\limits_{a=2}^{n} (\alpha_a+j_a )B(1+j_a,\alpha_a)}.\label{eq:highest_of_set_final_betas}
\end{gather}
This concludes our propensity derivation for general Beta distributions.
A consequence is that the iterative solution proposed by \citet{Miller2015} in Eq.~\ref{eq:incl-excl2} need not be solved iteratively, but Eq.~\ref{eq:highest_of_set_final_betas} can be applied directly instead.

Note that the solution implied by Eqs.~\ref{eq:incl-excl3} and~\ref{eq:lowest_of_set} has computational complexity $\mathcal{O}(\prod_{j=2}^{n}\alpha_j)$, the computational complexity of Eq.~\ref{eq:highest_of_set_final_betas} is $\mathcal{O}(\prod_{j=2}^{n}\beta_j)$.
In various practical use-cases, the empirical success rate of an action might be low (i.e. $\alpha<\beta$), implying that the former solution would be preferable.
However, it requires iteratively solving Eq.~\ref{eq:lowest_of_set} for all possible subsets, which is not accounted for in big-$\mathcal{O}$ notation but might prove to be cumbersome.
Practical implementations should consider these factors.
Finally, we note that logarithmic transformations can help avoid common numerical stability issues with the Beta function, as it involves factorials.
\section{Empirical Validation \& Discussion}
The main contributions of this article are theoretical in nature: we derive novel exact expressions for action propensities under various common practical TS scenarios, enabling the use of counterfactual estimators when either or both the logging and target policy operate under this common probabilistic action selection procedure.

Without our expressions, the only estimation approach for $\pi(a|x)$ would be Monte Carlo sampling, which is prohibitively expensive for the vast majority of practical applications.
We provide a proof-of-concept empirical validation along with a practical implementation of our methods, showcasing unbiased offline evaluation results using TS policies on the Open Bandit Pipeline~\cite{Saito2021_OBP}.

We leverage a similar experimental setup to \citet{Gupta2024}, where we simulate realistic synthetic data from a recommendation policy and wish to use it to:
\begin{enumerate*}
    \item train a Bayesian logistic regression model to be used in a TS policy $\pi_{\rm TS}$~\cite[Alg. 3]{Chapelle2011},
    \item validate that IPS-based off-policy estimators succeed in unbiasedly estimating the reward obtained when deploying $\pi_{\rm TS}$.
\end{enumerate*}
Further details on the experimental setup can be found in Appendix~\ref{app:exp}.

The results in Figure~\ref{fig:exp} validate that the action propensities we have derived can be directly applied to IPS-based estimators (traditional~\cite{Bottou2013}, self-normalised~\cite{Swaminathan2015}, and $\beta$-IPS~\cite{Gupta2024}), to obtain unbiased and consistent estimates for the average reward we would expect to obtain from deploying a TS-based stochastic policy.
The ground truth for $V(\pi_{\rm TS})$ is obtained as an empirical average over $10^{6}$ contexts and $10^{4}$ Monte Carlo samples, minimising sampling variance.

As hinted at earlier in this article---these results imply that the same propensities can be used to estimate any parameter of the counterfactual reward distribution induced by $\pi_{\rm TS}$~\cite{Chandak2021}, opening up a range of potential counterfactual inference analyses that would have previously been considered infeasible.

\begin{figure}[!t]
    \centering
    \includegraphics[width=\linewidth]{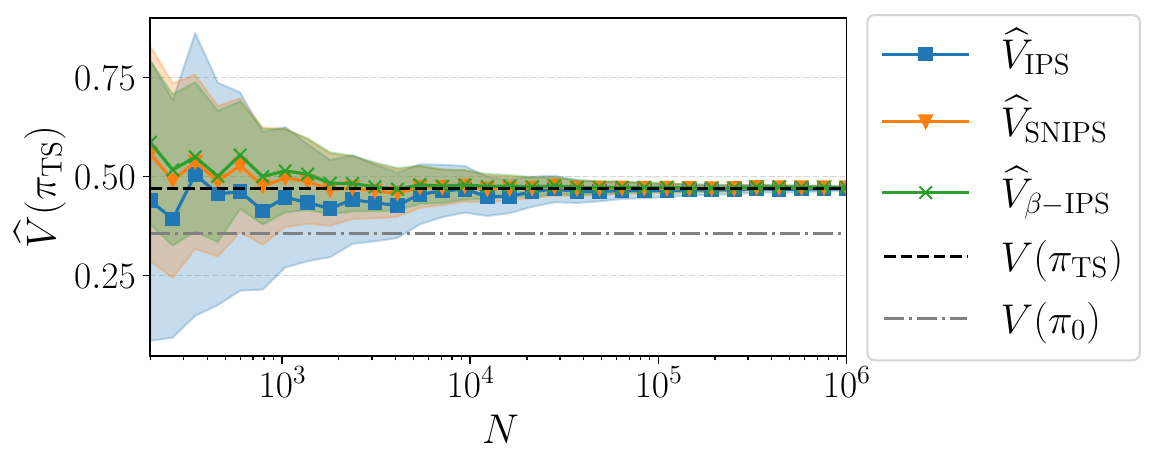}
    \caption{Action propensities obtained through our method can successfully be applied to consistently and unbiasedly estimate the expected value for a Thompson sampling policy $\pi_{\rm TS}$. As the sample size $N$ grows, all IPS-based estimators converge to the true value.}
    \label{fig:exp}
\end{figure}

To aid in the reproducibility of our work, all source code to reproduce these results as well as general implementations of our methods are made available under an open-source software license at \href{https://github.com/olivierjeunen/ope-ts-recsys-2025}{github.com/olivierjeunen/ope-ts-recsys-2025}.
\section{Conclusions \& Outlook}
Our work addresses a fundamental limitation in answering causal or counterfactual inference questions with policies that leverage Thompson sampling.
By deriving exact and efficiently computable expressions for action propensities under various common parameter and outcome distributions, we enable the use of off-policy estimators that rely on importance sampling ideas in these settings.
We show that this facilitates unbiased offline evaluation of recommendation policies using TS---through reproducible simulated experiments on the OBP~\cite{Saito2021_OBP}.
These results open up further avenues for future work that leverages off-policy estimators beyond traditional settings, broadening the situations where questions of causal and counterfactual inference can be answered reliably.

\begin{acks}
    The author would like to thank Duane Rich for valuable comments on earlier versions of this manuscript, and Marcelo Hartmann for insights regarding the case of (log)normal distributions.
\end{acks}

\bibliographystyle{ACM-Reference-Format}
\bibliography{bibliography}


\begin{thebibliography}{66}


\ifx \showCODEN    \undefined \def \showCODEN     #1{\unskip}     \fi
\ifx \showISBNx    \undefined \def \showISBNx     #1{\unskip}     \fi
\ifx \showISBNxiii \undefined \def \showISBNxiii  #1{\unskip}     \fi
\ifx \showISSN     \undefined \def \showISSN      #1{\unskip}     \fi
\ifx \showLCCN     \undefined \def \showLCCN      #1{\unskip}     \fi
\ifx \shownote     \undefined \def \shownote      #1{#1}          \fi
\ifx \showarticletitle \undefined \def \showarticletitle #1{#1}   \fi
\ifx \showURL      \undefined \def \showURL       {\relax}        \fi
\providecommand\bibfield[2]{#2}
\providecommand\bibinfo[2]{#2}
\providecommand\natexlab[1]{#1}
\providecommand\showeprint[2][]{arXiv:#2}

\bibitem[Amat et~al\mbox{.}(2018)]%
        {Amat2018}
\bibfield{author}{\bibinfo{person}{Fernando Amat}, \bibinfo{person}{Ashok Chandrashekar}, \bibinfo{person}{Tony Jebara}, {and} \bibinfo{person}{Justin Basilico}.} \bibinfo{year}{2018}\natexlab{}.
\newblock \showarticletitle{Artwork personalization at netflix}. In \bibinfo{booktitle}{\emph{Proc. of the 12th ACM Conference on Recommender Systems}} \emph{(\bibinfo{series}{RecSys '18})}. \bibinfo{publisher}{ACM}, \bibinfo{pages}{487–488}.
\newblock
\showISBNx{9781450359016}
\href{https://doi.org/10.1145/3240323.3241729}{doi:\nolinkurl{10.1145/3240323.3241729}}


\bibitem[Bendada et~al\mbox{.}(2020)]%
        {Bendada2020}
\bibfield{author}{\bibinfo{person}{Walid Bendada}, \bibinfo{person}{Guillaume Salha}, {and} \bibinfo{person}{Th\'{e}o Bontempelli}.} \bibinfo{year}{2020}\natexlab{}.
\newblock \showarticletitle{Carousel Personalization in Music Streaming Apps with Contextual Bandits}. In \bibinfo{booktitle}{\emph{Proc. of the 14th ACM Conference on Recommender Systems}} \emph{(\bibinfo{series}{RecSys '20})}. \bibinfo{publisher}{ACM}, \bibinfo{pages}{420–425}.
\newblock
\showISBNx{9781450375832}
\href{https://doi.org/10.1145/3383313.3412217}{doi:\nolinkurl{10.1145/3383313.3412217}}


\bibitem[Bottou et~al\mbox{.}(2013)]%
        {Bottou2013}
\bibfield{author}{\bibinfo{person}{L{{\'e}}on Bottou}, \bibinfo{person}{Jonas Peters}, \bibinfo{person}{Joaquin Qui{{\~n}}onero-Candela}, \bibinfo{person}{Denis~X. Charles}, \bibinfo{person}{D.~Max Chickering}, \bibinfo{person}{Elon Portugaly}, \bibinfo{person}{Dipankar Ray}, \bibinfo{person}{Patrice Simard}, {and} \bibinfo{person}{Ed Snelson}.} \bibinfo{year}{2013}\natexlab{}.
\newblock \showarticletitle{Counterfactual Reasoning and Learning Systems: The Example of Computational Advertising}.
\newblock \bibinfo{journal}{\emph{Journal of Machine Learning Research}} \bibinfo{volume}{14}, \bibinfo{number}{101} (\bibinfo{year}{2013}), \bibinfo{pages}{3207--3260}.
\newblock
\urldef\tempurl%
\url{http://jmlr.org/papers/v14/bottou13a.html}
\showURL{%
\tempurl}


\bibitem[Brod\'{e}n et~al\mbox{.}(2018)]%
        {Broden2018}
\bibfield{author}{\bibinfo{person}{Bj\"{o}rn Brod\'{e}n}, \bibinfo{person}{Mikael Hammar}, \bibinfo{person}{Bengt~J. Nilsson}, {and} \bibinfo{person}{Dimitris Paraschakis}.} \bibinfo{year}{2018}\natexlab{}.
\newblock \showarticletitle{Ensemble Recommendations via Thompson Sampling: an Experimental Study within e-Commerce}. In \bibinfo{booktitle}{\emph{Proc. of the 23rd International Conference on Intelligent User Interfaces}} \emph{(\bibinfo{series}{IUI '18})}. \bibinfo{publisher}{ACM}, \bibinfo{pages}{19–29}.
\newblock
\showISBNx{9781450349451}
\href{https://doi.org/10.1145/3172944.3172967}{doi:\nolinkurl{10.1145/3172944.3172967}}


\bibitem[Chandak et~al\mbox{.}(2021)]%
        {Chandak2021}
\bibfield{author}{\bibinfo{person}{Yash Chandak}, \bibinfo{person}{Scott Niekum}, \bibinfo{person}{Bruno da Silva}, \bibinfo{person}{Erik Learned-Miller}, \bibinfo{person}{Emma Brunskill}, {and} \bibinfo{person}{Philip~S. Thomas}.} \bibinfo{year}{2021}\natexlab{}.
\newblock \showarticletitle{Universal Off-Policy Evaluation}. In \bibinfo{booktitle}{\emph{Advances in Neural Information Processing Systems}}, Vol.~\bibinfo{volume}{34}. \bibinfo{publisher}{Curran Associates, Inc.}, \bibinfo{pages}{27475--27490}.
\newblock
\urldef\tempurl%
\url{https://proceedings.neurips.cc/paper_files/paper/2021/file/e71e5cd119bbc5797164fb0cd7fd94a4-Paper.pdf}
\showURL{%
\tempurl}


\bibitem[Chapelle and Li(2011)]%
        {Chapelle2011}
\bibfield{author}{\bibinfo{person}{Olivier Chapelle} {and} \bibinfo{person}{Lihong Li}.} \bibinfo{year}{2011}\natexlab{}.
\newblock \showarticletitle{An Empirical Evaluation of Thompson Sampling}. In \bibinfo{booktitle}{\emph{Advances in Neural Information Processing Systems}}, Vol.~\bibinfo{volume}{24}. \bibinfo{publisher}{Curran Associates, Inc.}
\newblock
\urldef\tempurl%
\url{https://proceedings.neurips.cc/paper/2011/file/e53a0a2978c28872a4505bdb51db06dc-Paper.pdf}
\showURL{%
\tempurl}


\bibitem[Chen et~al\mbox{.}(2019)]%
        {chen2019top}
\bibfield{author}{\bibinfo{person}{Minmin Chen}, \bibinfo{person}{Alex Beutel}, \bibinfo{person}{Paul Covington}, \bibinfo{person}{Sagar Jain}, \bibinfo{person}{Francois Belletti}, {and} \bibinfo{person}{Ed~H Chi}.} \bibinfo{year}{2019}\natexlab{}.
\newblock \showarticletitle{Top-k off-policy correction for a REINFORCE recommender system}. In \bibinfo{booktitle}{\emph{Proc. of the Twelfth ACM International Conference on Web Search and Data Mining}}. \bibinfo{pages}{456--464}.
\newblock


\bibitem[Dudík et~al\mbox{.}(2014)]%
        {Dudik2014}
\bibfield{author}{\bibinfo{person}{Miroslav Dudík}, \bibinfo{person}{Dumitru Erhan}, \bibinfo{person}{John Langford}, {and} \bibinfo{person}{Lihong Li}.} \bibinfo{year}{2014}\natexlab{}.
\newblock \showarticletitle{Doubly Robust Policy Evaluation and Optimization}.
\newblock \bibinfo{journal}{\emph{Statist. Sci.}} \bibinfo{volume}{29}, \bibinfo{number}{4} (\bibinfo{year}{2014}), \bibinfo{pages}{485--511}.
\newblock
\showISSN{08834237, 21688745}
\urldef\tempurl%
\url{http://www.jstor.org/stable/43288496}
\showURL{%
\tempurl}


\bibitem[Durrett(2019)]%
        {Durret2019}
\bibfield{author}{\bibinfo{person}{Rick Durrett}.} \bibinfo{year}{2019}\natexlab{}.
\newblock \bibinfo{booktitle}{\emph{Probability: Theory and Examples}}. Vol.~\bibinfo{volume}{5th Ed.}
\newblock \bibinfo{publisher}{Cambridge university press}.
\newblock


\bibitem[Eide et~al\mbox{.}(2022)]%
        {Eide2022}
\bibfield{author}{\bibinfo{person}{Simen Eide}, \bibinfo{person}{David~S. Leslie}, {and} \bibinfo{person}{Arnoldo Frigessi}.} \bibinfo{year}{2022}\natexlab{}.
\newblock \showarticletitle{Dynamic slate recommendation with gated recurrent units and Thompson sampling}.
\newblock \bibinfo{journal}{\emph{Data Mining and Knowledge Discovery}} \bibinfo{volume}{36}, \bibinfo{number}{5} (\bibinfo{date}{01 Sep} \bibinfo{year}{2022}), \bibinfo{pages}{1756--1786}.
\newblock
\showISSN{1573-756X}
\href{https://doi.org/10.1007/s10618-022-00849-w}{doi:\nolinkurl{10.1007/s10618-022-00849-w}}


\bibitem[Farajtabar et~al\mbox{.}(2018)]%
        {Farajtabar2018}
\bibfield{author}{\bibinfo{person}{Mehrdad Farajtabar}, \bibinfo{person}{Yinlam Chow}, {and} \bibinfo{person}{Mohammad Ghavamzadeh}.} \bibinfo{year}{2018}\natexlab{}.
\newblock \showarticletitle{More Robust Doubly Robust Off-policy Evaluation}. In \bibinfo{booktitle}{\emph{Proc. of the 35th International Conference on Machine Learning}} \emph{(\bibinfo{series}{ICML'18}, Vol.~\bibinfo{volume}{80})}. \bibinfo{publisher}{PMLR}, \bibinfo{pages}{1447--1456}.
\newblock


\bibitem[Gao et~al\mbox{.}(2024)]%
        {Gao2024}
\bibfield{author}{\bibinfo{person}{Chen Gao}, \bibinfo{person}{Yu Zheng}, \bibinfo{person}{Wenjie Wang}, \bibinfo{person}{Fuli Feng}, \bibinfo{person}{Xiangnan He}, {and} \bibinfo{person}{Yong Li}.} \bibinfo{year}{2024}\natexlab{}.
\newblock \showarticletitle{Causal Inference in Recommender Systems: A Survey and Future Directions}.
\newblock \bibinfo{journal}{\emph{ACM Trans. Inf. Syst.}} \bibinfo{volume}{42}, \bibinfo{number}{4}, Article \bibinfo{articleno}{88} (\bibinfo{date}{Feb.} \bibinfo{year}{2024}), \bibinfo{numpages}{32}~pages.
\newblock
\showISSN{1046-8188}
\href{https://doi.org/10.1145/3639048}{doi:\nolinkurl{10.1145/3639048}}


\bibitem[Gilotte et~al\mbox{.}(2018)]%
        {Gilotte2018}
\bibfield{author}{\bibinfo{person}{Alexandre Gilotte}, \bibinfo{person}{Cl\'{e}ment Calauz\`{e}nes}, \bibinfo{person}{Thomas Nedelec}, \bibinfo{person}{Alexandre Abraham}, {and} \bibinfo{person}{Simon Doll\'{e}}.} \bibinfo{year}{2018}\natexlab{}.
\newblock \showarticletitle{Offline {A/B} Testing for Recommender Systems}. In \bibinfo{booktitle}{\emph{Proc. of the Eleventh ACM International Conference on Web Search and Data Mining}} \emph{(\bibinfo{series}{WSDM '18})}. \bibinfo{publisher}{ACM}, \bibinfo{pages}{198–206}.
\newblock
\showISBNx{9781450355810}
\urldef\tempurl%
\url{https://doi.org/10.1145/3159652.3159687}
\showURL{%
\tempurl}


\bibitem[Gruson et~al\mbox{.}(2019)]%
        {Gruson2019}
\bibfield{author}{\bibinfo{person}{Alois Gruson}, \bibinfo{person}{Praveen Chandar}, \bibinfo{person}{Christophe Charbuillet}, \bibinfo{person}{James McInerney}, \bibinfo{person}{Samantha Hansen}, \bibinfo{person}{Damien Tardieu}, {and} \bibinfo{person}{Ben Carterette}.} \bibinfo{year}{2019}\natexlab{}.
\newblock \showarticletitle{Offline Evaluation to Make Decisions About Playlist Recommendation Algorithms}. In \bibinfo{booktitle}{\emph{Proc. of the Twelfth ACM International Conference on Web Search and Data Mining}} \emph{(\bibinfo{series}{WSDM '19})}. \bibinfo{publisher}{ACM}, \bibinfo{pages}{420–428}.
\newblock
\showISBNx{9781450359405}
\href{https://doi.org/10.1145/3289600.3291027}{doi:\nolinkurl{10.1145/3289600.3291027}}


\bibitem[Gupta et~al\mbox{.}(2024a)]%
        {Gupta2024_WSDM}
\bibfield{author}{\bibinfo{person}{Shashank Gupta}, \bibinfo{person}{Philipp Hager}, \bibinfo{person}{Jin Huang}, \bibinfo{person}{Ali Vardasbi}, {and} \bibinfo{person}{Harrie Oosterhuis}.} \bibinfo{year}{2024}\natexlab{a}.
\newblock \showarticletitle{Unbiased Learning to Rank: On Recent Advances and Practical Applications}. In \bibinfo{booktitle}{\emph{Proc. of the 17th ACM International Conference on Web Search and Data Mining}} \emph{(\bibinfo{series}{WSDM '24})}. \bibinfo{publisher}{ACM}, \bibinfo{pages}{1118–1121}.
\newblock
\showISBNx{9798400703713}
\href{https://doi.org/10.1145/3616855.3636451}{doi:\nolinkurl{10.1145/3616855.3636451}}


\bibitem[Gupta et~al\mbox{.}(2024b)]%
        {Gupta2024}
\bibfield{author}{\bibinfo{person}{Shashank Gupta}, \bibinfo{person}{Olivier Jeunen}, \bibinfo{person}{Harrie Oosterhuis}, {and} \bibinfo{person}{Maarten de Rijke}.} \bibinfo{year}{2024}\natexlab{b}.
\newblock \showarticletitle{Optimal Baseline Corrections for Off-Policy Contextual Bandits}. In \bibinfo{booktitle}{\emph{Proc. of the 18th ACM Conference on Recommender Systems}} \emph{(\bibinfo{series}{RecSys '24})}. \bibinfo{publisher}{ACM}, \bibinfo{pages}{722–732}.
\newblock
\showISBNx{9798400705052}
\href{https://doi.org/10.1145/3640457.3688105}{doi:\nolinkurl{10.1145/3640457.3688105}}


\bibitem[Gupta et~al\mbox{.}(2024c)]%
        {Gupta2024_CIKM}
\bibfield{author}{\bibinfo{person}{Shashank Gupta}, \bibinfo{person}{Harrie Oosterhuis}, {and} \bibinfo{person}{Maarten de Rijke}.} \bibinfo{year}{2024}\natexlab{c}.
\newblock \showarticletitle{Practical and Robust Safety Guarantees for Advanced Counterfactual Learning to Rank}. In \bibinfo{booktitle}{\emph{Proc. of the 33rd ACM International Conference on Information and Knowledge Management}} \emph{(\bibinfo{series}{CIKM '24})}. \bibinfo{publisher}{ACM}, \bibinfo{pages}{737–747}.
\newblock
\showISBNx{9798400704369}
\href{https://doi.org/10.1145/3627673.3679531}{doi:\nolinkurl{10.1145/3627673.3679531}}


\bibitem[Hartmann(2017)]%
        {Hartmann2017}
\bibfield{author}{\bibinfo{person}{Marcelo Hartmann}.} \bibinfo{year}{2017}\natexlab{}.
\newblock \bibinfo{title}{Extending Owen's integral table and a new multivariate Bernoulli distribution}.
\newblock
\showeprint[arxiv]{1704.04736}~[stat.ME]
\urldef\tempurl%
\url{https://arxiv.org/abs/1704.04736}
\showURL{%
\tempurl}


\bibitem[Ionides(2008)]%
        {Ionides2008}
\bibfield{author}{\bibinfo{person}{Edward~L. Ionides}.} \bibinfo{year}{2008}\natexlab{}.
\newblock \showarticletitle{Truncated Importance Sampling}.
\newblock \bibinfo{journal}{\emph{Journal of Computational and Graphical Statistics}} \bibinfo{volume}{17}, \bibinfo{number}{2} (\bibinfo{year}{2008}), \bibinfo{pages}{295--311}.
\newblock


\bibitem[Jeunen(2019)]%
        {Jeunen2019}
\bibfield{author}{\bibinfo{person}{Olivier Jeunen}.} \bibinfo{year}{2019}\natexlab{}.
\newblock \showarticletitle{Revisiting Offline Evaluation for Implicit-Feedback Recommender Systems}. In \bibinfo{booktitle}{\emph{Proc. of the 13th ACM Conference on Recommender Systems}} \emph{(\bibinfo{series}{RecSys '19})}. \bibinfo{publisher}{ACM}, \bibinfo{pages}{596–600}.
\newblock
\showISBNx{9781450362436}
\href{https://doi.org/10.1145/3298689.3347069}{doi:\nolinkurl{10.1145/3298689.3347069}}


\bibitem[Jeunen(2021)]%
        {Jeunen2021Thesis}
\bibfield{author}{\bibinfo{person}{Olivier Jeunen}.} \bibinfo{year}{2021}\natexlab{}.
\newblock \emph{\bibinfo{title}{Offline Approaches to Recommendation with Online Success}}.
\newblock \bibinfo{thesistype}{Ph.\,D. Dissertation}. \bibinfo{school}{University of Antwerp}.
\newblock


\bibitem[Jeunen(2023)]%
        {Jeunen2023_C3PO}
\bibfield{author}{\bibinfo{person}{Olivier Jeunen}.} \bibinfo{year}{2023}\natexlab{}.
\newblock \showarticletitle{A Probabilistic Position Bias Model for Short-Video Recommendation Feeds}. In \bibinfo{booktitle}{\emph{Proc. of the 17th ACM Conference on Recommender Systems}} \emph{(\bibinfo{series}{RecSys '23})}. \bibinfo{publisher}{ACM}, \bibinfo{pages}{675–681}.
\newblock
\showISBNx{9798400702419}
\href{https://doi.org/10.1145/3604915.3608777}{doi:\nolinkurl{10.1145/3604915.3608777}}


\bibitem[Jeunen and Goethals(2021a)]%
        {Jeunen2021_Pessimism}
\bibfield{author}{\bibinfo{person}{Olivier Jeunen} {and} \bibinfo{person}{Bart Goethals}.} \bibinfo{year}{2021}\natexlab{a}.
\newblock \showarticletitle{Pessimistic Reward Models for Off-Policy Learning in Recommendation}. In \bibinfo{booktitle}{\emph{Proc. of the 15th ACM Conference on Recommender Systems}} \emph{(\bibinfo{series}{RecSys '21})}. \bibinfo{publisher}{ACM}, \bibinfo{pages}{63–74}.
\newblock
\showISBNx{9781450384582}
\href{https://doi.org/10.1145/3460231.3474247}{doi:\nolinkurl{10.1145/3460231.3474247}}


\bibitem[Jeunen and Goethals(2021b)]%
        {Jeunen2021_TS}
\bibfield{author}{\bibinfo{person}{Olivier Jeunen} {and} \bibinfo{person}{Bart Goethals}.} \bibinfo{year}{2021}\natexlab{b}.
\newblock \showarticletitle{Top-K Contextual Bandits with Equity of Exposure}. In \bibinfo{booktitle}{\emph{Proc. of the 15th ACM Conference on Recommender Systems}} \emph{(\bibinfo{series}{RecSys '21})}. \bibinfo{publisher}{ACM}, \bibinfo{pages}{310–320}.
\newblock
\showISBNx{9781450384582}
\href{https://doi.org/10.1145/3460231.3474248}{doi:\nolinkurl{10.1145/3460231.3474248}}


\bibitem[Jeunen and Goethals(2023)]%
        {Jeunen2023}
\bibfield{author}{\bibinfo{person}{Olivier Jeunen} {and} \bibinfo{person}{Bart Goethals}.} \bibinfo{year}{2023}\natexlab{}.
\newblock \showarticletitle{Pessimistic Decision-Making for Recommender Systems}.
\newblock \bibinfo{journal}{\emph{ACM Trans. Recomm. Syst.}} \bibinfo{volume}{1}, \bibinfo{number}{1}, Article \bibinfo{articleno}{4} (\bibinfo{date}{feb} \bibinfo{year}{2023}), \bibinfo{numpages}{27}~pages.
\newblock
\href{https://doi.org/10.1145/3568029}{doi:\nolinkurl{10.1145/3568029}}


\bibitem[Jeunen et~al\mbox{.}(2022)]%
        {CONSEQUENCES2022}
\bibfield{author}{\bibinfo{person}{Olivier Jeunen}, \bibinfo{person}{Thorsten Joachims}, \bibinfo{person}{Harrie Oosterhuis}, \bibinfo{person}{Yuta Saito}, {and} \bibinfo{person}{Flavian Vasile}.} \bibinfo{year}{2022}\natexlab{}.
\newblock \showarticletitle{CONSEQUENCES — Causality, Counterfactuals and Sequential Decision-Making for Recommender Systems}. In \bibinfo{booktitle}{\emph{Proc. of the 16th ACM Conference on Recommender Systems}} \emph{(\bibinfo{series}{RecSys '22})}. \bibinfo{publisher}{ACM}, \bibinfo{pages}{654–657}.
\newblock
\showISBNx{9781450392785}
\href{https://doi.org/10.1145/3523227.3547409}{doi:\nolinkurl{10.1145/3523227.3547409}}


\bibitem[Jeunen et~al\mbox{.}(2024a)]%
        {Jeunen2024Multi}
\bibfield{author}{\bibinfo{person}{Olivier Jeunen}, \bibinfo{person}{Jatin Mandav}, \bibinfo{person}{Ivan Potapov}, \bibinfo{person}{Nakul Agarwal}, \bibinfo{person}{Sourabh Vaid}, \bibinfo{person}{Wenzhe Shi}, {and} \bibinfo{person}{Aleksei Ustimenko}.} \bibinfo{year}{2024}\natexlab{a}.
\newblock \showarticletitle{Multi-Objective Recommendation via Multivariate Policy Learning}. In \bibinfo{booktitle}{\emph{Proc. of the 18th ACM Conference on Recommender Systems}} \emph{(\bibinfo{series}{RecSys '24})}. \bibinfo{publisher}{ACM}, \bibinfo{pages}{712–721}.
\newblock
\showISBNx{9798400705052}
\href{https://doi.org/10.1145/3640457.3688132}{doi:\nolinkurl{10.1145/3640457.3688132}}


\bibitem[Jeunen et~al\mbox{.}(2023)]%
        {Jeunen2023_AuctionGym}
\bibfield{author}{\bibinfo{person}{Olivier Jeunen}, \bibinfo{person}{Sean Murphy}, {and} \bibinfo{person}{Ben Allison}.} \bibinfo{year}{2023}\natexlab{}.
\newblock \showarticletitle{Off-Policy Learning-to-Bid with AuctionGym}. In \bibinfo{booktitle}{\emph{Proc. of the 29th ACM SIGKDD Conference on Knowledge Discovery and Data Mining}} \emph{(\bibinfo{series}{KDD '23})}. \bibinfo{publisher}{ACM}, \bibinfo{pages}{4219–4228}.
\newblock
\showISBNx{9798400701030}
\href{https://doi.org/10.1145/3580305.3599877}{doi:\nolinkurl{10.1145/3580305.3599877}}


\bibitem[Jeunen et~al\mbox{.}(2024b)]%
        {Jeunen2023_nDCG}
\bibfield{author}{\bibinfo{person}{Olivier Jeunen}, \bibinfo{person}{Ivan Potapov}, {and} \bibinfo{person}{Aleksei Ustimenko}.} \bibinfo{year}{2024}\natexlab{b}.
\newblock \showarticletitle{On (Normalised) Discounted Cumulative Gain as an Off-Policy Evaluation Metric for Top-n Recommendation}. In \bibinfo{booktitle}{\emph{Proc. of the 30th ACM SIGKDD Conference on Knowledge Discovery and Data Mining}} \emph{(\bibinfo{series}{KDD '24})}. \bibinfo{publisher}{ACM}, \bibinfo{pages}{1222–1233}.
\newblock
\showISBNx{9798400704901}
\href{https://doi.org/10.1145/3637528.3671687}{doi:\nolinkurl{10.1145/3637528.3671687}}


\bibitem[Jeunen et~al\mbox{.}(2020)]%
        {Jeunen2020}
\bibfield{author}{\bibinfo{person}{Olivier Jeunen}, \bibinfo{person}{David Rohde}, \bibinfo{person}{Flavian Vasile}, {and} \bibinfo{person}{Martin Bompaire}.} \bibinfo{year}{2020}\natexlab{}.
\newblock \showarticletitle{Joint Policy-Value Learning for Recommendation}. In \bibinfo{booktitle}{\emph{Proc. of the 26th ACM SIGKDD International Conference on Knowledge Discovery \& Data Mining}} \emph{(\bibinfo{series}{KDD '20})}. \bibinfo{publisher}{ACM}, \bibinfo{pages}{1223–1233}.
\newblock
\showISBNx{9781450379984}
\href{https://doi.org/10.1145/3394486.3403175}{doi:\nolinkurl{10.1145/3394486.3403175}}


\bibitem[Jeunen and Ustimenko(2024)]%
        {Jeunen2024_DeltaOPE}
\bibfield{author}{\bibinfo{person}{Olivier Jeunen} {and} \bibinfo{person}{Aleksei Ustimenko}.} \bibinfo{year}{2024}\natexlab{}.
\newblock \showarticletitle{$\Delta$-OPE: Off-Policy Estimation with Pairs of Policies}. In \bibinfo{booktitle}{\emph{Proc. of the 18th ACM Conference on Recommender Systems}} \emph{(\bibinfo{series}{RecSys '24})}. \bibinfo{publisher}{ACM}, \bibinfo{pages}{878–883}.
\newblock
\showISBNx{9798400705052}
\href{https://doi.org/10.1145/3640457.3688162}{doi:\nolinkurl{10.1145/3640457.3688162}}


\bibitem[Jin et~al\mbox{.}(2023)]%
        {Jin2023}
\bibfield{author}{\bibinfo{person}{Tianyuan Jin}, \bibinfo{person}{Xianglin Yang}, \bibinfo{person}{Xiaokui Xiao}, {and} \bibinfo{person}{Pan Xu}.} \bibinfo{year}{2023}\natexlab{}.
\newblock \showarticletitle{Thompson Sampling with Less Exploration is Fast and Optimal}. In \bibinfo{booktitle}{\emph{Proc. of the 40th International Conference on Machine Learning}} \emph{(\bibinfo{series}{Proc. of Machine Learning Research}, Vol.~\bibinfo{volume}{202})}. \bibinfo{publisher}{PMLR}, \bibinfo{pages}{15239--15261}.
\newblock
\urldef\tempurl%
\url{https://proceedings.mlr.press/v202/jin23b.html}
\showURL{%
\tempurl}


\bibitem[Joachims et~al\mbox{.}(2021)]%
        {Joachims2021}
\bibfield{author}{\bibinfo{person}{Thorsten Joachims}, \bibinfo{person}{Ben London}, \bibinfo{person}{Yi Su}, \bibinfo{person}{Adith Swaminathan}, {and} \bibinfo{person}{Lequn Wang}.} \bibinfo{year}{2021}\natexlab{}.
\newblock \showarticletitle{Recommendations as Treatments}.
\newblock \bibinfo{journal}{\emph{AI Magazine}} \bibinfo{volume}{42}, \bibinfo{number}{3} (\bibinfo{date}{Nov.} \bibinfo{year}{2021}), \bibinfo{pages}{19--30}.
\newblock
\href{https://doi.org/10.1609/aimag.v42i3.18141}{doi:\nolinkurl{10.1609/aimag.v42i3.18141}}


\bibitem[Joachims et~al\mbox{.}(2017)]%
        {Joachims2017}
\bibfield{author}{\bibinfo{person}{Thorsten Joachims}, \bibinfo{person}{Adith Swaminathan}, {and} \bibinfo{person}{Tobias Schnabel}.} \bibinfo{year}{2017}\natexlab{}.
\newblock \showarticletitle{Unbiased Learning-to-Rank with Biased Feedback}. In \bibinfo{booktitle}{\emph{Proc. of the Tenth ACM International Conference on Web Search and Data Mining}} \emph{(\bibinfo{series}{WSDM '17})}. \bibinfo{publisher}{ACM}, \bibinfo{pages}{781–789}.
\newblock
\showISBNx{9781450346757}
\href{https://doi.org/10.1145/3018661.3018699}{doi:\nolinkurl{10.1145/3018661.3018699}}


\bibitem[Kaufmann et~al\mbox{.}(2012)]%
        {Kaufmann2012}
\bibfield{author}{\bibinfo{person}{Emilie Kaufmann}, \bibinfo{person}{Olivier Cappe}, {and} \bibinfo{person}{Aurelien Garivier}.} \bibinfo{year}{2012}\natexlab{}.
\newblock \showarticletitle{On Bayesian Upper Confidence Bounds for Bandit Problems}. In \bibinfo{booktitle}{\emph{Proc. of the Fifteenth International Conference on Artificial Intelligence and Statistics}} \emph{(\bibinfo{series}{ICML '12}, Vol.~\bibinfo{volume}{22})}. \bibinfo{publisher}{PMLR}, \bibinfo{pages}{592--600}.
\newblock
\urldef\tempurl%
\url{https://proceedings.mlr.press/v22/kaufmann12.html}
\showURL{%
\tempurl}


\bibitem[Lattimore and Szepesv{\'a}ri(2020)]%
        {lattimore2020bandit}
\bibfield{author}{\bibinfo{person}{Tor Lattimore} {and} \bibinfo{person}{Csaba Szepesv{\'a}ri}.} \bibinfo{year}{2020}\natexlab{}.
\newblock \bibinfo{booktitle}{\emph{Bandit algorithms}}.
\newblock \bibinfo{publisher}{Cambridge University Press}.
\newblock


\bibitem[Li et~al\mbox{.}(2011)]%
        {Li2011}
\bibfield{author}{\bibinfo{person}{Lihong Li}, \bibinfo{person}{Wei Chu}, \bibinfo{person}{John Langford}, {and} \bibinfo{person}{Xuanhui Wang}.} \bibinfo{year}{2011}\natexlab{}.
\newblock \showarticletitle{Unbiased offline evaluation of contextual-bandit-based news article recommendation algorithms}. In \bibinfo{booktitle}{\emph{Proc. of the Fourth ACM International Conference on Web Search and Data Mining}} \emph{(\bibinfo{series}{WSDM '11})}. \bibinfo{publisher}{ACM}, \bibinfo{pages}{297–306}.
\newblock
\showISBNx{9781450304931}
\href{https://doi.org/10.1145/1935826.1935878}{doi:\nolinkurl{10.1145/1935826.1935878}}


\bibitem[Lu et~al\mbox{.}(2018)]%
        {Lu2018}
\bibfield{author}{\bibinfo{person}{Xiuyuan Lu}, \bibinfo{person}{Zheng Wen}, {and} \bibinfo{person}{Branislav Kveton}.} \bibinfo{year}{2018}\natexlab{}.
\newblock \showarticletitle{Efficient online recommendation via low-rank ensemble sampling}. In \bibinfo{booktitle}{\emph{Proc. of the 12th ACM Conference on Recommender Systems}} \emph{(\bibinfo{series}{RecSys '18})}. \bibinfo{publisher}{ACM}, \bibinfo{pages}{460–464}.
\newblock
\showISBNx{9781450359016}
\href{https://doi.org/10.1145/3240323.3240408}{doi:\nolinkurl{10.1145/3240323.3240408}}


\bibitem[Ma et~al\mbox{.}(2020)]%
        {ma2020off}
\bibfield{author}{\bibinfo{person}{Jiaqi Ma}, \bibinfo{person}{Zhe Zhao}, \bibinfo{person}{Xinyang Yi}, \bibinfo{person}{Ji Yang}, \bibinfo{person}{Minmin Chen}, \bibinfo{person}{Jiaxi Tang}, \bibinfo{person}{Lichan Hong}, {and} \bibinfo{person}{Ed~H Chi}.} \bibinfo{year}{2020}\natexlab{}.
\newblock \showarticletitle{Off-policy learning in two-stage recommender systems}. In \bibinfo{booktitle}{\emph{Proc. of The Web Conference 2020}}. \bibinfo{pages}{463--473}.
\newblock


\bibitem[Miller(2015)]%
        {Miller2015}
\bibfield{author}{\bibinfo{person}{Evan Miller}.} \bibinfo{year}{2015}\natexlab{}.
\newblock \bibinfo{title}{Formulas for Bayesian A/B testing}.
\newblock
\urldef\tempurl%
\url{https://www.evanmiller.org/bayesian-ab-testing.html}
\showURL{%
\tempurl}


\bibitem[Murphy(2022)]%
        {Murphy2022}
\bibfield{author}{\bibinfo{person}{Kevin~P. Murphy}.} \bibinfo{year}{2022}\natexlab{}.
\newblock \bibinfo{booktitle}{\emph{Probabilistic Machine Learning: An introduction}}.
\newblock \bibinfo{publisher}{MIT Press}.
\newblock
\urldef\tempurl%
\url{http://probml.github.io/book1}
\showURL{%
\tempurl}


\bibitem[Oosterhuis and de~Rijke(2020)]%
        {Oosterhuis2020}
\bibfield{author}{\bibinfo{person}{Harrie Oosterhuis} {and} \bibinfo{person}{Maarten de Rijke}.} \bibinfo{year}{2020}\natexlab{}.
\newblock \showarticletitle{Policy-Aware Unbiased Learning to Rank for Top-k Rankings}. In \bibinfo{booktitle}{\emph{Proc. of the 43rd International ACM SIGIR Conference on Research and Development in Information Retrieval}} \emph{(\bibinfo{series}{SIGIR '20})}. \bibinfo{publisher}{ACM}, \bibinfo{pages}{489–498}.
\newblock
\showISBNx{9781450380164}
\href{https://doi.org/10.1145/3397271.3401102}{doi:\nolinkurl{10.1145/3397271.3401102}}


\bibitem[Owen(2013)]%
        {Owen2013}
\bibfield{author}{\bibinfo{person}{Art~B. Owen}.} \bibinfo{year}{2013}\natexlab{}.
\newblock \bibinfo{booktitle}{\emph{Monte Carlo theory, methods and examples}}.
\newblock


\bibitem[Robbins(1952)]%
        {robbins1952some}
\bibfield{author}{\bibinfo{person}{Herbert Robbins}.} \bibinfo{year}{1952}\natexlab{}.
\newblock \showarticletitle{Some aspects of the sequential design of experiments}.
\newblock \bibinfo{journal}{\emph{Bull. Amer. Math. Soc.}}  \bibinfo{volume}{58} (\bibinfo{year}{1952}), \bibinfo{pages}{527--535}.
\newblock


\bibitem[Russo et~al\mbox{.}(2018)]%
        {Russo2018}
\bibfield{author}{\bibinfo{person}{Daniel~J. Russo}, \bibinfo{person}{Benjamin~Van Roy}, \bibinfo{person}{Abbas Kazerouni}, \bibinfo{person}{Ian Osband}, {and} \bibinfo{person}{Zheng Wen}.} \bibinfo{year}{2018}\natexlab{}.
\newblock \showarticletitle{A Tutorial on Thompson Sampling}.
\newblock \bibinfo{journal}{\emph{Foundations and Trends® in Machine Learning}} \bibinfo{volume}{11}, \bibinfo{number}{1} (\bibinfo{year}{2018}), \bibinfo{pages}{1--96}.
\newblock
\showISSN{1935-8237}
\href{https://doi.org/10.1561/2200000070}{doi:\nolinkurl{10.1561/2200000070}}


\bibitem[Sagtani et~al\mbox{.}(2024)]%
        {Sagtani2024}
\bibfield{author}{\bibinfo{person}{Hitesh Sagtani}, \bibinfo{person}{Madan~Gopal Jhawar}, \bibinfo{person}{Rishabh Mehrotra}, {and} \bibinfo{person}{Olivier Jeunen}.} \bibinfo{year}{2024}\natexlab{}.
\newblock \showarticletitle{Ad-load Balancing via Off-policy Learning in a Content Marketplace}. In \bibinfo{booktitle}{\emph{Proc. of the 17th ACM International Conference on Web Search and Data Mining}} \emph{(\bibinfo{series}{WSDM '24})}. \bibinfo{publisher}{ACM}, \bibinfo{pages}{586–595}.
\newblock
\showISBNx{9798400703713}
\href{https://doi.org/10.1145/3616855.3635846}{doi:\nolinkurl{10.1145/3616855.3635846}}


\bibitem[Saito et~al\mbox{.}(2021)]%
        {Saito2021_OBP}
\bibfield{author}{\bibinfo{person}{Yuta Saito}, \bibinfo{person}{Shunsuke Aihara}, \bibinfo{person}{Megumi Matsutani}, {and} \bibinfo{person}{Yusuke Narita}.} \bibinfo{year}{2021}\natexlab{}.
\newblock \showarticletitle{Open Bandit Dataset and Pipeline: Towards Realistic and Reproducible Off-Policy Evaluation}. In \bibinfo{booktitle}{\emph{Proc. of the Neural Information Processing Systems Track on Datasets and Benchmarks}}, Vol.~\bibinfo{volume}{1}.
\newblock
\urldef\tempurl%
\url{https://datasets-benchmarks-proceedings.neurips.cc/paper_files/paper/2021/file/33e75ff09dd601bbe69f351039152189-Paper-round2.pdf}
\showURL{%
\tempurl}


\bibitem[Saito and Joachims(2021)]%
        {Saito2021}
\bibfield{author}{\bibinfo{person}{Yuta Saito} {and} \bibinfo{person}{Thorsten Joachims}.} \bibinfo{year}{2021}\natexlab{}.
\newblock \showarticletitle{Counterfactual Learning and Evaluation for Recommender Systems: Foundations, Implementations, and Recent Advances}. In \bibinfo{booktitle}{\emph{Proc. of the 15th ACM Conference on Recommender Systems}} \emph{(\bibinfo{series}{RecSys '21})}. \bibinfo{publisher}{ACM}, \bibinfo{pages}{828–830}.
\newblock
\showISBNx{9781450384582}
\href{https://doi.org/10.1145/3460231.3473320}{doi:\nolinkurl{10.1145/3460231.3473320}}


\bibitem[Saito and Joachims(2022)]%
        {Saito2022_MIPS}
\bibfield{author}{\bibinfo{person}{Yuta Saito} {and} \bibinfo{person}{Thorsten Joachims}.} \bibinfo{year}{2022}\natexlab{}.
\newblock \showarticletitle{Off-Policy Evaluation for Large Action Spaces via Embeddings}. In \bibinfo{booktitle}{\emph{Proc. of the 39th International Conference on Machine Learning}} \emph{(\bibinfo{series}{Proc. of Machine Learning Research}, Vol.~\bibinfo{volume}{162})}. \bibinfo{publisher}{PMLR}, \bibinfo{pages}{19089--19122}.
\newblock
\urldef\tempurl%
\url{https://proceedings.mlr.press/v162/saito22a.html}
\showURL{%
\tempurl}


\bibitem[Saito et~al\mbox{.}(2023)]%
        {Saito2023}
\bibfield{author}{\bibinfo{person}{Yuta Saito}, \bibinfo{person}{Qingyang Ren}, {and} \bibinfo{person}{Thorsten Joachims}.} \bibinfo{year}{2023}\natexlab{}.
\newblock \showarticletitle{Off-Policy Evaluation for Large Action Spaces via Conjunct Effect Modeling}. In \bibinfo{booktitle}{\emph{Proc. of the 40th International Conference on Machine Learning}} \emph{(\bibinfo{series}{Proc. of Machine Learning Research}, Vol.~\bibinfo{volume}{202})}. \bibinfo{publisher}{PMLR}, \bibinfo{pages}{29734--29759}.
\newblock
\urldef\tempurl%
\url{https://proceedings.mlr.press/v202/saito23b.html}
\showURL{%
\tempurl}


\bibitem[Sakhi et~al\mbox{.}(2023)]%
        {Sakhi2023}
\bibfield{author}{\bibinfo{person}{Otmane Sakhi}, \bibinfo{person}{Pierre Alquier}, {and} \bibinfo{person}{Nicolas Chopin}.} \bibinfo{year}{2023}\natexlab{}.
\newblock \showarticletitle{{PAC}-{B}ayesian Offline Contextual Bandits With Guarantees}. In \bibinfo{booktitle}{\emph{Proc. of the 40th International Conference on Machine Learning}} \emph{(\bibinfo{series}{Proceedings of Machine Learning Research}, Vol.~\bibinfo{volume}{202})}, \bibfield{editor}{\bibinfo{person}{Andreas Krause}, \bibinfo{person}{Emma Brunskill}, \bibinfo{person}{Kyunghyun Cho}, \bibinfo{person}{Barbara Engelhardt}, \bibinfo{person}{Sivan Sabato}, {and} \bibinfo{person}{Jonathan Scarlett}} (Eds.). \bibinfo{publisher}{PMLR}, \bibinfo{pages}{29777--29799}.
\newblock
\urldef\tempurl%
\url{https://proceedings.mlr.press/v202/sakhi23a.html}
\showURL{%
\tempurl}


\bibitem[Su and Chen(2023)]%
        {Su2023}
\bibfield{author}{\bibinfo{person}{Yi Su} {and} \bibinfo{person}{Minmin Chen}.} \bibinfo{year}{2023}\natexlab{}.
\newblock \showarticletitle{Nonlinear Bandits Exploration for Recommendations}. In \bibinfo{booktitle}{\emph{Proc. of the 17th ACM Conference on Recommender Systems}} \emph{(\bibinfo{series}{RecSys '23})}. \bibinfo{publisher}{ACM}, \bibinfo{pages}{1054–1057}.
\newblock
\showISBNx{9798400702419}
\href{https://doi.org/10.1145/3604915.3610245}{doi:\nolinkurl{10.1145/3604915.3610245}}


\bibitem[Su et~al\mbox{.}(2020)]%
        {Su2020}
\bibfield{author}{\bibinfo{person}{Yi Su}, \bibinfo{person}{Maria Dimakopoulou}, \bibinfo{person}{Akshay Krishnamurthy}, {and} \bibinfo{person}{Miroslav Dudik}.} \bibinfo{year}{2020}\natexlab{}.
\newblock \showarticletitle{Doubly robust off-policy evaluation with shrinkage}. In \bibinfo{booktitle}{\emph{Proc. of the 37th International Conference on Machine Learning}} \emph{(\bibinfo{series}{Proc. of Machine Learning Research}, Vol.~\bibinfo{volume}{119})}. \bibinfo{publisher}{PMLR}, \bibinfo{pages}{9167--9176}.
\newblock
\urldef\tempurl%
\url{https://proceedings.mlr.press/v119/su20a.html}
\showURL{%
\tempurl}


\bibitem[Su et~al\mbox{.}(2024)]%
        {Su2024}
\bibfield{author}{\bibinfo{person}{Yi Su}, \bibinfo{person}{Haokai Lu}, \bibinfo{person}{Yuening Li}, \bibinfo{person}{Liang Liu}, \bibinfo{person}{Shuchao Bi}, \bibinfo{person}{Ed~H. Chi}, {and} \bibinfo{person}{Minmin Chen}.} \bibinfo{year}{2024}\natexlab{}.
\newblock \showarticletitle{Multi-Task Neural Linear Bandit for Exploration in Recommender Systems}. In \bibinfo{booktitle}{\emph{Proc. of the 30th ACM SIGKDD Conference on Knowledge Discovery and Data Mining}} \emph{(\bibinfo{series}{KDD '24})}. \bibinfo{publisher}{ACM}, \bibinfo{pages}{5723–5730}.
\newblock
\showISBNx{9798400704901}
\href{https://doi.org/10.1145/3637528.3671649}{doi:\nolinkurl{10.1145/3637528.3671649}}


\bibitem[Swaminathan and Joachims(2015)]%
        {Swaminathan2015}
\bibfield{author}{\bibinfo{person}{Adith Swaminathan} {and} \bibinfo{person}{Thorsten Joachims}.} \bibinfo{year}{2015}\natexlab{}.
\newblock \showarticletitle{The Self-Normalized Estimator for Counterfactual Learning}. In \bibinfo{booktitle}{\emph{Advances in Neural Information Processing Systems}}, Vol.~\bibinfo{volume}{28}. \bibinfo{publisher}{Curran Associates, Inc.}
\newblock
\urldef\tempurl%
\url{https://proceedings.neurips.cc/paper_files/paper/2015/file/39027dfad5138c9ca0c474d71db915c3-Paper.pdf}
\showURL{%
\tempurl}


\bibitem[Thompson(1933)]%
        {Thompson1933}
\bibfield{author}{\bibinfo{person}{William~R. Thompson}.} \bibinfo{year}{1933}\natexlab{}.
\newblock \showarticletitle{On the Likelihood that One Unknown Probability Exceeds Another in View of the Evidence of Two Samples}.
\newblock \bibinfo{journal}{\emph{Biometrika}} \bibinfo{volume}{25}, \bibinfo{number}{3/4} (\bibinfo{year}{1933}), \bibinfo{pages}{285--294}.
\newblock
\showISSN{00063444}
\urldef\tempurl%
\url{http://www.jstor.org/stable/2332286}
\showURL{%
\tempurl}


\bibitem[van~den Akker et~al\mbox{.}(2024)]%
        {vandenAkker2024}
\bibfield{author}{\bibinfo{person}{Bram van~den Akker}, \bibinfo{person}{Olivier Jeunen}, \bibinfo{person}{Ying Li}, \bibinfo{person}{Ben London}, \bibinfo{person}{Zahra Nazari}, {and} \bibinfo{person}{Devesh Parekh}.} \bibinfo{year}{2024}\natexlab{}.
\newblock \showarticletitle{Practical Bandits: An Industry Perspective}. In \bibinfo{booktitle}{\emph{Proc. of the 17th ACM International Conference on Web Search and Data Mining}} \emph{(\bibinfo{series}{WSDM '24})}. \bibinfo{publisher}{ACM}, \bibinfo{pages}{1132–1135}.
\newblock
\showISBNx{9798400703713}
\href{https://doi.org/10.1145/3616855.3636449}{doi:\nolinkurl{10.1145/3616855.3636449}}


\bibitem[Vasile et~al\mbox{.}(2020)]%
        {Vasile2020}
\bibfield{author}{\bibinfo{person}{Flavian Vasile}, \bibinfo{person}{David Rohde}, \bibinfo{person}{Olivier Jeunen}, {and} \bibinfo{person}{Amine Benhalloum}.} \bibinfo{year}{2020}\natexlab{}.
\newblock \showarticletitle{A Gentle Introduction to Recommendation as Counterfactual Policy Learning}. In \bibinfo{booktitle}{\emph{Proc. of the 28th ACM Conference on User Modeling, Adaptation and Personalization}} \emph{(\bibinfo{series}{UMAP '20})}. \bibinfo{publisher}{ACM}, \bibinfo{pages}{392–393}.
\newblock
\showISBNx{9781450368612}
\href{https://doi.org/10.1145/3340631.3398666}{doi:\nolinkurl{10.1145/3340631.3398666}}


\bibitem[Virtanen~et al.(2020)]%
        {Virtanen2020}
\bibfield{author}{\bibinfo{person}{Pauli Virtanen~et al.}} \bibinfo{year}{2020}\natexlab{}.
\newblock \showarticletitle{SciPy 1.0: fundamental algorithms for scientific computing in Python}.
\newblock \bibinfo{journal}{\emph{Nature Methods}} \bibinfo{volume}{17}, \bibinfo{number}{3} (\bibinfo{date}{01 Mar} \bibinfo{year}{2020}), \bibinfo{pages}{261--272}.
\newblock
\showISSN{1548-7105}
\href{https://doi.org/10.1038/s41592-019-0686-2}{doi:\nolinkurl{10.1038/s41592-019-0686-2}}


\bibitem[Vlassis et~al\mbox{.}(2019)]%
        {Vlassis2019}
\bibfield{author}{\bibinfo{person}{Nikos Vlassis}, \bibinfo{person}{Aurelien Bibaut}, \bibinfo{person}{Maria Dimakopoulou}, {and} \bibinfo{person}{Tony Jebara}.} \bibinfo{year}{2019}\natexlab{}.
\newblock \showarticletitle{On the Design of Estimators for Bandit Off-Policy Evaluation}. In \bibinfo{booktitle}{\emph{Proc. of the 36th International Conference on Machine Learning}} \emph{(\bibinfo{series}{ICML'19}, Vol.~\bibinfo{volume}{97})}. \bibinfo{publisher}{PMLR}, \bibinfo{pages}{6468--6476}.
\newblock


\bibitem[Wang et~al\mbox{.}(2020)]%
        {Wang2020_Causal}
\bibfield{author}{\bibinfo{person}{Yixin Wang}, \bibinfo{person}{Dawen Liang}, \bibinfo{person}{Laurent Charlin}, {and} \bibinfo{person}{David~M. Blei}.} \bibinfo{year}{2020}\natexlab{}.
\newblock \showarticletitle{Causal Inference for Recommender Systems}. In \bibinfo{booktitle}{\emph{Proc. of the 14th ACM Conference on Recommender Systems}} \emph{(\bibinfo{series}{RecSys '20})}. \bibinfo{publisher}{ACM}, \bibinfo{pages}{426–431}.
\newblock
\showISBNx{9781450375832}
\href{https://doi.org/10.1145/3383313.3412225}{doi:\nolinkurl{10.1145/3383313.3412225}}


\bibitem[Watson et~al\mbox{.}(2021)]%
        {Watson2021}
\bibfield{author}{\bibinfo{person}{Joe Watson}, \bibinfo{person}{Jihao Andreas~Lin}, \bibinfo{person}{Pascal Klink}, \bibinfo{person}{Joni Pajarinen}, {and} \bibinfo{person}{Jan Peters}.} \bibinfo{year}{2021}\natexlab{}.
\newblock \showarticletitle{Latent Derivative Bayesian Last Layer Networks}. In \bibinfo{booktitle}{\emph{Proc. of The 24th International Conference on Artificial Intelligence and Statistics}} \emph{(\bibinfo{series}{Proc. of Machine Learning Research}, Vol.~\bibinfo{volume}{130})}. \bibinfo{publisher}{PMLR}, \bibinfo{pages}{1198--1206}.
\newblock
\urldef\tempurl%
\url{https://proceedings.mlr.press/v130/watson21a.html}
\showURL{%
\tempurl}


\bibitem[Yang et~al\mbox{.}(2018)]%
        {Yang2018}
\bibfield{author}{\bibinfo{person}{Longqi Yang}, \bibinfo{person}{Yin Cui}, \bibinfo{person}{Yuan Xuan}, \bibinfo{person}{Chenyang Wang}, \bibinfo{person}{Serge Belongie}, {and} \bibinfo{person}{Deborah Estrin}.} \bibinfo{year}{2018}\natexlab{}.
\newblock \showarticletitle{Unbiased Offline Recommender Evaluation for Missing-Not-at-Random Implicit Feedback}. In \bibinfo{booktitle}{\emph{Proc. of the 12th ACM Conference on Recommender Systems}} \emph{(\bibinfo{series}{RecSys '18})}. \bibinfo{publisher}{ACM}, \bibinfo{pages}{279–287}.
\newblock
\showISBNx{9781450359016}
\href{https://doi.org/10.1145/3240323.3240355}{doi:\nolinkurl{10.1145/3240323.3240355}}


\bibitem[Zhang et~al\mbox{.}(2017)]%
        {Zhang2017}
\bibfield{author}{\bibinfo{person}{Liangpeng Zhang}, \bibinfo{person}{Ke Tang}, {and} \bibinfo{person}{Xin Yao}.} \bibinfo{year}{2017}\natexlab{}.
\newblock \showarticletitle{Log-normality and Skewness of Estimated State/Action Values in Reinforcement Learning}. In \bibinfo{booktitle}{\emph{Advances in Neural Information Processing Systems}}, \bibfield{editor}{\bibinfo{person}{I.~Guyon}, \bibinfo{person}{U.~Von Luxburg}, \bibinfo{person}{S.~Bengio}, \bibinfo{person}{H.~Wallach}, \bibinfo{person}{R.~Fergus}, \bibinfo{person}{S.~Vishwanathan}, {and} \bibinfo{person}{R.~Garnett}} (Eds.), Vol.~\bibinfo{volume}{30}. \bibinfo{publisher}{Curran Associates, Inc.}
\newblock
\urldef\tempurl%
\url{https://proceedings.neurips.cc/paper_files/paper/2017/file/69a5b5995110b36a9a347898d97a610e-Paper.pdf}
\showURL{%
\tempurl}


\bibitem[Zhang et~al\mbox{.}(2021)]%
        {Zhang2021}
\bibfield{author}{\bibinfo{person}{Weitong Zhang}, \bibinfo{person}{Dongruo Zhou}, \bibinfo{person}{Lihong Li}, {and} \bibinfo{person}{Quanquan Gu}.} \bibinfo{year}{2021}\natexlab{}.
\newblock \showarticletitle{Neural Thompson Sampling}. In \bibinfo{booktitle}{\emph{International Conference on Learning Representations}} \emph{(\bibinfo{series}{ICLR '21})}.
\newblock
\urldef\tempurl%
\url{https://openreview.net/forum?id=tkAtoZkcUnm}
\showURL{%
\tempurl}


\bibitem[Zhu and Van~Roy(2023)]%
        {Zhu2023}
\bibfield{author}{\bibinfo{person}{Zheqing Zhu} {and} \bibinfo{person}{Benjamin Van~Roy}.} \bibinfo{year}{2023}\natexlab{}.
\newblock \showarticletitle{Deep Exploration for Recommendation Systems}. In \bibinfo{booktitle}{\emph{Proc. of the 17th ACM Conference on Recommender Systems}} \emph{(\bibinfo{series}{RecSys '23})}. \bibinfo{publisher}{ACM}, \bibinfo{pages}{963–970}.
\newblock
\showISBNx{9798400702419}
\href{https://doi.org/10.1145/3604915.3608855}{doi:\nolinkurl{10.1145/3604915.3608855}}


\end{thebibliography}

\newpage
\appendix
\onecolumn
\section{\citet{Miller2015}'s Derivation}\label{app:miller}
Using the definition of the Beta distribution's probability density function and subsequently evaluating the inner integral, we can write:
\begin{align}
&\mathsf{P}(p_i > p_j) = \int_0^1 \int_{p_j}^1 \frac{p_j^{\alpha_j-1}(1-p_j)^{\beta_j-1}}{B(\alpha_j, \beta_j)} \frac{{p_i}^{\alpha_i-1}(1-p_i)^{\beta_i-1}}{B(\alpha_i, \beta_i)} {\rm d}p_i {\rm d}p_j = 1 - \int_0^1 \frac{p_j^{\alpha_j-1}(1-p_j)^{\beta_j-1}}{B(\alpha_j,\beta_j)}I_{p_j}(\alpha_i, \beta_i){\rm d}p_j.
\end{align}

When assuming $a$ and $b$ are natural numbers, the regularised incomplete beta function $I_{x}(a,b)$ can equivalently be expressed as:
\begin{equation}
I_x(a, b) = 1 - \sum_{i=0}^{a-1} \frac{x^{i}(1 - x)^b}{(b+i)B(1+i,b)}.
\end{equation}

Rewriting then yields:
\begin{align}
\mathsf{P}(p_i > p_j) &= 1 - \int_0^1 \frac{p_j^{\alpha_j-1}(1-p_j)^{\beta_j-1}}{B(\alpha_j,\beta_j)} \left(1 - \sum_{k=0}^{\alpha_i-1}{\frac{p_j^k(1-p_j)^{\beta_i}}{(\beta_i+k)B(1+k, \beta_i)}}\right){\rm d}p_j \nonumber\\
&= 1 - 1 + \int_0^1 \frac{p_j^{\alpha_j-1}(1-p_j)^{\beta_j-1}}{B(\alpha_j,\beta_j)} \sum_{k=0}^{\alpha_i-1}{\frac{p_j^k(1-p_j)^{\beta_i}}{(\beta_i+k)B(1+k, \beta_i)}}{\rm d}p_j \nonumber\\
&= \int_0^1 \sum_{k=0}^{\alpha_i-1}{\frac{p_j^{\alpha_j-1+k}(1-p_j)^{\beta_j+\beta_i-1}}{(\beta_i+k)B(\alpha_j, \beta_j)B(1+k, \beta_i)}}{\rm d}p_j \nonumber\\
&=\sum_{k=0}^{\alpha_i-1}\int_0^1{\frac{p_j^{\alpha_j-1+k}(1-p_j)^{\beta_j+\beta_i-1}}{(\beta_i+k)B(\alpha_j, \beta_j)B(1+k, \beta_i)}}{\rm d}p_j \nonumber\\
&= \sum_{k=0}^{\alpha_i-1}\frac{B(\alpha_j+k,\beta_j+\beta_i)}{(\beta_i+k)B(\alpha_j, \beta_j)B(1+k, \beta_i)} \int_0^1{\frac{p_j^{\alpha_j-1+k}(1-p_j)^{\beta_j+\beta_i-1}}{B(\alpha_j+k,\beta_j+\beta_i)}}{\rm d}p_j \nonumber \\
&= \sum_{k=0}^{\alpha_i-1}\frac{B(\alpha_j+k,\beta_j+\beta_i)}{(\beta_i+k)B(\alpha_j, \beta_j)B(1+k, \beta_i)}.
\end{align}


\section{Experimental Setup} \label{app:exp}
For the illustrative purposes of this empirical validation, and to ensure that these results can be reproduced efficiently, we set the number of latent dimensions in the context embeddings to $10$, the size of the action space to $10$, and we consider a softmax logging policy $\pi_{0}$ with temperature $\beta=-1.5$, using the standard synthetic setup for the Open Bandit Pipeline~\cite{Saito2021_OBP}.
We set the prior on the weights in the Bayesian logistic regression model to $\mathcal{N}(0;1e3)$ and train the model on $2^{11}$ samples obtained through $\pi_{0}$.
For the induced policy $\pi_{\rm TS}$, we aim to estimate the reward we would obtain when deploying it $V(\pi_{\rm TS})$, from a dataset $\mathcal{D}$ collected under $\pi_{0}$.
We vary the dataset size $|\mathcal{D}| \in [10^{2},10^6]$ and visualise 99\% confidence intervals for various counterfactual estimators in Figure~\ref{fig:exp}.
As theory suggests, the resulting estimators are consistent and unbiased, implying that they converge to the true value $V(\pi_{\rm TS})$ as more data is used by the estimators.

All source code necessary to reproduce these results is available at  \href{https://github.com/olivierjeunen/ope-ts-recsys-2025}{github.com/olivierjeunen/ope-ts-recsys-2025}.

\end{document}